%% file: d2d-cc_ICC2019.tex
\newif\ifdoublecol

\doublecoltrue
\documentclass[conference]{IEEEtran}

\IEEEoverridecommandlockouts

\usepackage{etex} 
\usepackage{algorithm}

\usepackage[cmex10]{amsmath}

\ifCLASSINFOpdf
  \usepackage[pdftex]{graphicx}
  \usepackage{epstopdf}
\else
  \usepackage[dvips]{graphicx}
  \graphicspath{{../eps/}}
  \DeclareGraphicsExtensions{.eps}
\fi
\usepackage{cite}

\usepackage[cmex10]{amsmath}
\usepackage{amssymb}
\usepackage{enumerate}
\usepackage{algorithm}
\usepackage{algpseudocode}
\usepackage{mdwtab}
\usepackage{amssymb}
\usepackage{eqparbox}
\usepackage{amsfonts}
\usepackage{bm}
\usepackage{booktabs}
\usepackage{bbm}
\usepackage[caption=false,font=footnotesize]{subfig}
\usepackage{url}
\usepackage{acro}
\usepackage{bbding}
\usepackage{comment}
\usepackage{tikz}
\usepackage{xcolor}
\usetikzlibrary{decorations.pathreplacing}

\input{commands}

\input{abbrev}

\newcommand{\ds}{\displaystyle}

\newcommand{\M}[1]{\mathbf{#1}}

\begin{document}

\title{D2D Assisted Beamforming for Coded Caching}

\author{\IEEEauthorblockN{Hamidreza Bakhshzad Mahmoodi$^\star$, Jarkko Kaleva$^\star$, Seyed Pooya Shariatpanahi$^*$, Babak Khalaj$^\dagger$ and Antti T\"olli$^\star$}
\IEEEauthorblockA{
    $\star$ Centre for Wireless Communications, University of Oulu, P.O. Box 4500, 90014, Finland \\
    $*$ Institute for Research in Fundamental Sciences (IPM) Tehran, Iran, \
    $\dagger$ Sharif University of Technology, Tehran, Iran \\
    \textrm{firstname.lastname@oulu.fi, pooya@ipm.ir, khalaj@sharif.edu}
    }
    	\thanks{This work was supported by the Academy of Finland under grants no. 319059 (Coded Collaborative Caching for Wireless Energy Efficiency) and 318927 (6Genesis Flagship).}

}  
    
\maketitle

\input{abstract}

\IEEEpeerreviewmaketitle

\input{intro}

\input{sysmodel}

\input{examples}
\input{general}
\input{simres}


\input{conclusions}



\bibliographystyle{IEEEtran}
\bibliography{IEEEabrv,conf_short,jour_short,references}

\end{document}

%% file: commands.tex
\newcommand{\herm}{^{\mbox{\scriptsize H}}}

\newcommand{\vbar}{\raisebox{.17ex}{\rule{.04em}{1.35ex}}}
\newcommand{\vbarind}{\raisebox{.01ex}{\rule{.04em}{1.1ex}}}
\newcommand{\R}{\ifmmode{\rm I}\hspace{-.2em}{\rm R} \else ${\rm I}\hspace{-.2em}{\rm R}$ \fi}
\newcommand{\T}{\ifmmode{\rm I}\hspace{-.2em}{\rm T} \else ${\rm I}\hspace{-.2em}{\rm T}$ \fi}
\newcommand{\N}{\ifmmode{\rm I}\hspace{-.2em}{\rm N} \else \mbox{${\rm I}\hspace{-.2em}{\rm N}$} \fi}
\newcommand{\B}{\ifmmode{\rm I}\hspace{-.2em}{\rm B} \else \mbox{${\rm I}\hspace{-.2em}{\rm B}$} \fi}
\newcommand{\Hil}{\ifmmode{\rm I}\hspace{-.2em}{\rm H} \else \mbox{${\rm I}\hspace{-.2em}{\rm H}$} \fi}
\newcommand{\C}{\ifmmode\hspace{.2em}\vbar\hspace{-.31em}{\rm C} \else \mbox{$\hspace{.2em}\vbar\hspace{-.31em}{\rm C}$} \fi}
\newcommand{\Cind}{\ifmmode\hspace{.2em}\vbarind\hspace{-.25em}{\rm C} \else \mbox{$\hspace{.2em}\vbarind\hspace{-.25em}{\rm C}$} \fi}
\newcommand{\Q}{\ifmmode\hspace{.2em}\vbar\hspace{-.31em}{\rm Q} \else \mbox{$\hspace{.2em}\vbar\hspace{-.31em}{\rm Q}$} \fi}
\newcommand{\Z}{\ifmmode{\rm Z}\hspace{-.28em}{\rm Z} \else ${\rm Z}\hspace{-.28em}{\rm Z}$ \fi}



%% file: abbrev.tex
\DeclareAcronym{ADMM}{
    short = ADMM,
    long = alternating direction method of multipliers,
    list = Alternating Direction Method of Multipliers,
    class = abbrev
}

\DeclareAcronym{AoA}{
    short = AoA,
    long = angle-of-arrival,
    list = Angle-of-Arrival,
    class = abbrev
}

\DeclareAcronym{AoD}{
    short = AoD,
    long = angle-of-departure,
    list = Angle-of-Departure,
    class = abbrev
}

\DeclareAcronym{BB}{
    short = BB,
    long = base band,
    list = Base Band,
    class = abbrev
}

\DeclareAcronym{BC}{
    short = BC,
    long = broadcast channel,
    list = Broadcast Channel,
    class = abbrev
}

\DeclareAcronym{BS}{
    short = BS,
    long = base station,
    list = Base Station,
    class = abbrev
}

\DeclareAcronym{BR}{
    short = BR,
    long = best response,
    list = Best Response, 
    class = abbrev
}

\DeclareAcronym{CB}{
    short = CB,
    long = coordinated beamforming,
    list = Coordinated Beamforming,
    class = abbrev
}

\DeclareAcronym{CC}{
    short = CC,
    long = coded caching,
    list = Coded Caching,
    class = abbrev
}

\DeclareAcronym{CE}{
    short = CE,
    long = channel estimation,
    list = Channel Estimation,
    class = abbrev
}

\DeclareAcronym{CoMP}{
    short = CoMP,
    long = coordinated multi-point transmission,
    list = Coordinated Multi-Point Transmission,
    class = abbrev
}

\DeclareAcronym{CRAN}{
    short = C-RAN,
    long = cloud radio access network,
    list = Cloud Radio Access Network,
    class = abbrev
}

\DeclareAcronym{CSE}{
    short = CSE,
    long = channel specific estimation,
    list = Channel Specific Estimation,
    class = abbrev
}

\DeclareAcronym{CSI}{
    short = CSI,
    long = channel state information,
    list = Channel State Information,
    class = abbrev
}

\DeclareAcronym{CSIT}{
    short = CSIT,
    long = channel state information at the transmitter,
    list = Channel State Information at the Transmitter,
    class = abbrev
}

\DeclareAcronym{CU}{
    short = CU,
    long = central unit,
    list = Central Unit,
    class = abbrev
}

\DeclareAcronym{D2D}{
    short = D2D,
    long = device-to-device,
    list = Device-to-Device,
    class = abbrev
}

\DeclareAcronym{DE-ADMM}{
    short = DE-ADMM,
    long = direct estimation with alternating direction method of multipliers,
    list = Direct Estimation with Alternating Direction Method of Multipliers,
    class = abbrev
}

\DeclareAcronym{DE-BR}{
    short = DE-BR,
    long = direct estimation with best response,
    list = Direct Estimation with Best Response,
    class = abbrev
}

\DeclareAcronym{DE-SG}{
    short = DE-SG,
    long = direct estimation with stochastic gradient,
    list = Direct Estimation with Stochastic Gradient,
    class = abbrev
}

\DeclareAcronym{DFT}{
	short = DFT,
	long = discrete fourier transform,
	list = Discrete Fourier Transform,
	class = abbrev
}

\DeclareAcronym{DoF}{
    short = DoF,
    long = degrees of freedom,
    list = Degrees of Freedom,
    class = abbrev
}

\DeclareAcronym{DL}{
    short = DL,
    long = downlink,
    list = Downlink,
    class = abbrev
}

\DeclareAcronym{GD}{
	short = GD, 
	long = gradient descent,
	list = Gradeitn Descent,
	class = abbrev
}

\DeclareAcronym{IBC}{
    short = IBC,
    long = interfering broadcast channel,
    list = Interfering Broadcast Channel,
    class = abbrev
}

\DeclareAcronym{i.i.d.}{
    short = i.i.d.,
    long = independent and identically distributed,
    list = Independent and Identically Distributed,
    class = abbrev
}

\DeclareAcronym{JP}{
    short = JP,
    long = joint processing,
    list = Joint Processing,
    class = abbrev
}

\DeclareAcronym{KKT}{
    short = KKT,
    long = Karush-Kuhn-Tucker,
    class = abbrev
}

\DeclareAcronym{LOS}{
	short = LOS,
	long = line-of-sight,
	list = Line-of-Sight,
	class = abbrev
}

\DeclareAcronym{LS}{
    short = LS,
    long = least squares,
    list = Least Squares,
    class = abbrev
}

\DeclareAcronym{LTE}{
    short = LTE,
    long = Long Term Evolution,
    class = abbrev
}

\DeclareAcronym{LTE-A}{
    short = LTE-A,
    long = Long Term Evolution Advanced,
    class = abbrev
}

\DeclareAcronym{MIMO}{
    short = MIMO,
    long = multiple-input multiple-output,
    list = Multiple-Input Multiple-Output,
    class = abbrev
}

\DeclareAcronym{MISO}{
    short = MISO,
    long = multiple-input single-output,
    list = Multiple-Input Single-Output,
    class = abbrev
}

\DeclareAcronym{MSE}{
    short = MSE,
    long = mean-squared error,
    list = Mean-Squared Error,
    class = abbrev
}

\DeclareAcronym{MMSE}{
    short = MMSE,
    long = minimum mean-squared error,
    list = Minimum Mean-Squared Error,
    class = abbrev
}

\DeclareAcronym{mmWave}{
	short = mmWave,
	long = millimeter wave,
	list = Millimeter Wave,
	class = abbrev
}

\DeclareAcronym{MU-MIMO}{
    short = MU-MIMO,
    long = multi-user \ac{MIMO},
    list = Multi-User \ac{MIMO},
    class = abbrev
}

\DeclareAcronym{OTA}{
    short = OTA,
    long = over-the-air,
    list = Over-the-Air,
    class = abbrev
}

\DeclareAcronym{PSD}{
    short = PSD,
    long = positive semidefinite,
    list = Positive Semidefinite,
    class = abbrev
}

\DeclareAcronym{QoS}{
	short = QoS,
	long = quality of service,
	list = Quality of Service,
	class = abbrev
}

\DeclareAcronym{RCP}{
	short = RCP,
	long = remote central processor,
	list = Remote Central Processor,
	class = abbrev
}

\DeclareAcronym{RRH}{
    short = RRH,
    long = remote radio head,
    list = Remote Radio Head,
    class = abbrev
}

\DeclareAcronym{RSSI}{
    short = RSSI,
    long = received signal strength indicator,
    list = Received Signal Strength Indicator,
    class = abbrev
}

\DeclareAcronym{RX}{
	short = RX,
	long = receiver,
	list = Receiver,
	class = abbrev
}

\DeclareAcronym{SCA}{
    short = SCA,
    long = successive convex approximation,
    list = Successive Convex Approximation,
    class = abbrev
}

\DeclareAcronym{SG}{
    short = SG,
    long = stochastic gradient,
    list = Stochastic Gradient,
    class = abbrev
}

\DeclareAcronym{SNR}{
    short = SNR,
    long = signal-to-noise ratio,
    list = Signal-to-Noise Ratio,
    class = abbrev
}

\DeclareAcronym{SINR}{
    short = SINR,
    long = signal-to-interference-plus-noise ratio,
    list = Signal-to-Interference-plus-Noise Ratio,
    class = abbrev
}

\DeclareAcronym{SOCP}{
	short = SOCP, 
	long = second order cone program,
	list = Second Order Cone Program,
	class = abbrev
}

\DeclareAcronym{SSE}{
    short = SSE,
    long = stream specific estimation,
    list = Stream Specific Estimation,
    class = abbrev
}

\DeclareAcronym{SVD}{
	short = SVD,
	long = singular value decomposition,
	list = Singular Value Decomposition,
	class = abbrev
}

\DeclareAcronym{TDD}{
	short = TDD,
	long = time division duplex,
	list = Time Division Duplex,
	class = abbrev
}

\DeclareAcronym{TX}{
	short = TX,
	long = transmitter,
	list = Transmitter,
	class = abbrev
}

\DeclareAcronym{UE}{
    short = UE,
    long = user equipment,
    list = User Equipment,
    class = abbrev
}

\DeclareAcronym{UL}{
    short = UL,
    long = uplink,
    list = Uplink,
    class = abbrev
}

\DeclareAcronym{ULA}{
	short = ULA,
	long = uniform linear array,
	list = Uniform Linear Array,
	class = abbrev
}

\DeclareAcronym{UPA}{
    short = UPA,
    long = uniform planar array,
    list = Uniform Planar Array,
    class = abbrev
}

\DeclareAcronym{WMMSE}{
    short = WMMSE,
    long = weighted minimum mean-squared error,
    list = Weighted Minimum Mean-Squared Error,
    class = abbrev
}

\DeclareAcronym{WMSEMin}{
    short = WMSEMin,
    long = weighted sum \ac{MSE} minimization,
    list = Weighted sum \ac{MSE} Minimization,
    class = abbrev
}

\DeclareAcronym{WBAN}{
	short = WBAN,
	long = wireless body area network,
	list = Wireless Body Area Network,
	class = abbrev
}

\DeclareAcronym{WSRMax}{
    short = WSRMax,
    long = weighted sum rate maximization,
    list = Weighted Sum Rate Maximization,
    class = abbrev
}

%% file: abstract.tex

\begin{abstract}

Device-to-device (D2D) aided beamforming for coded caching is considered in finite signal-to-noise ratio regime. A novel beamforming scheme is proposed where the local cache content exchange among nearby users is exploited. The transmission is split into two phases: local D2D content exchange and downlink transmission. In the D2D phase, users can autonomously share content with the adjacent users. The downlink phase utilizes multicast beamforming to simultaneously serve all users to fulfill the remaining content requests. We first explain the main procedure via two simple examples and then present the general formulation. Furthermore, D2D transmission scenarios and conditions useful for minimizing the overall delivery time are identified. We also investigate the benefits of using D2D transmission for decreasing the transceiver complexity of multicast beamforming. By exploiting the direct D2D exchange of file fragments, the common multicasting rate for delivering the remaining file fragments in the downlink phase is increased providing greatly enhanced overall content delivery performance.

\end{abstract}

%% file: intro.tex

\section{Introduction}
\label{sec:intro}

Caching popular content near end-users is a widely accepted solution for supporting high quality content delivery in next generation networks. This solution includes benefiting from  off-peak hours of the network to move the content closer to the end-users, which will be used to mitigate the content delivery burden in network peak hours. Many recent papers have investigated the potentials of this paradigm to improve wireless networks performance, such as \cite{Shanmugam2016}, \cite{Bastug2014}, and \cite{Paschos2016}. A promising scheme in this context is proposed in \cite{MaddahAli-2014}, which is known as the so-called \emph{\Ac{CC}}  approach. In this scheme, instead of locally caching the entire files at the end-user, fragments of all files in the library are stored in all the users' cache memories. In the delivery phase, carefully formed coded messages are multicast to groups of users, which results in \emph{global caching gain}~\cite{MaddahAli-2014}.

\Ac{CC} has been shown to be greatly beneficial for both wired and wireless content delivery, under various assumptions~\cite{MaddahAli-2014, Karamchandi2016, Pedarsani2016, Shariatpanahi2016, Naderalizadeh2017,Naderalizadeh2017-2}. 
The original coded caching setup is extended in~\cite{Shariatpanahi2016}  to a multiple server scenario under different network 
topologies, aiming to further minimize the required delivery time of requested content. 
For high \ac{SNR} regime,~\cite{Naderalizadeh2017, Naderalizadeh2017-2, Zhang2017, Lampiris2018} show that coded caching can boost the performance of the wireless network in terms of Degrees-of-Freedom (DoF). Specifically, in wireless broadcast channels with a multiple-antenna base station, the global coded caching gain and the spatial multiplexing gain are shown to be additive and will increase the network data rate \cite{Shariatpanahi2016, Naderalizadeh2017, Lampiris2018}.  

In order to bridge the gap between high-SNR analysis of \ac{CC} and the practical finite-SNR scenarios, recent works on finite \ac{SNR} regime have also shown \ac{CC} to be greatly beneficial when the interference is properly accounted for~\cite{Ngo2017, Piovano2017, Shariatpanahi2017, Toll1806:Multicast,Tolli-Shariatpanahi-Kaleva-Khalaj-Arxiv18}. While, the works \cite{Ngo2017} and \cite{Piovano2017} use a rate-splitting approach to benefit from the global caching gain and the spatial multiplexing gain at finite SNR, the work \cite{Shariatpanahi2017} follows a Zero-Forcing (ZF) based approach (extending the ideas in \cite{Shariatpanahi2016} to the finite-SNR setup), which is also order-optimal in terms of DoF. Moreover, the work \cite{Toll1806:Multicast,Tolli-Shariatpanahi-Kaleva-Khalaj-Arxiv18} extends \cite{Shariatpanahi2017} to a general beamformer solution which manages the interaction between interference and noise in an optimal manner. The general interference management framework proposed in \cite{Toll1806:Multicast,Tolli-Shariatpanahi-Kaleva-Khalaj-Arxiv18}, improves the finite-SNR performance of the coded caching in wireless networks significantly. Moreover, the complexity issues associated with the corresponding optimization problem are addressed in \cite{Tolli-Shariatpanahi-Kaleva-Khalaj-Arxiv18}.

This paper considers a delivery scheme optimized for finite \ac{SNR} region, where the multicast beamforming~\cite{Tolli-Shariatpanahi-Kaleva-Khalaj-Arxiv18} of file fragments is complemented by allowing direct \ac{D2D} exchange of local cache content. 
Finding the optimal D2D opportunities in finite SNR is particularly challenging due to the high computational complexity for the DL multicast beamformer design. The optimal D2D/DL mode selection requires exhaustive search for D2D opportunities over a group of users, which quickly becomes computationally intractable. To over come these practical limitations, we provide a low complexity mode selection algorithm, which allows efficient determination of D2D opportunities even for large number users. The computational complexity of the proposed algorithm is greatly reduced with respect to the exhaustive search baseline while retaining comparable performance.


%% file: sysmodel.tex

\section{System Model}
\label{sec:sysmodel}

We consider a system consisting of a single $L$ antenna \ac{BS} and $K$ single antenna users. The \ac{BS} has a library of $N$ files, namely $\mathcal{W}=\{W_1,\dots,W_N\}$, where each file has the size of $F$ bits. The normalized cache size (memory) at each user is $M$ files. Each user $k$ caches a function of the files, denoted by $Z_k(W_1,\dots,W_N)$, which is stored in the {\it cache content placement} phase during off peak hours.  At the {\it content delivery phase}, user $k \in \{1,\dots K\}$ makes a request for the file $W_{d_k}, d_k \in [1 : N]$.  
\begin{figure}
    \centering 
    \includegraphics[width=0.8\columnwidth,keepaspectratio]{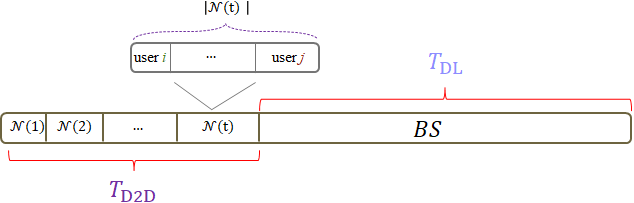}
    \caption{Time division in D2D assisted transmission. Total time needed to transmit all fragments of files to the users is $T_{\mathrm{D2D}}+T_{\mathrm{DL}}$.}
\label{fig:timeDiv}
\end{figure}

Upon the requests arrival, first we have a \ac{D2D} sub-phase which is divided into a number of D2D time slots. In each time slot $t$, a group of nearby users, denoted by set $\mathcal{N}(t)$, are instructed by the BS to locally exchange data (see Fig. \ref{fig:timeDiv}). Furthermore, each D2D time slot is divided into  $ |\mathcal{N}(t)|$ individual D2D transmissions. In each D2D transmission a user $i \in \mathcal{N}(t)$ transmits a coded message comprised of $\frac{1}{{KM}/{N}}$ of some file fragments denoted by $X^\mathrm{D2D}_i$ to an intended set of receivers $\mathcal{R}^{\mathcal{N}}(i) \subseteq \mathcal{N}(t)$, which are interested in decoding $X^\mathrm{D2D}_i$. Thus, the message $X^\mathrm{D2D}_i$ can be transmitted at rate\footnote{In this paper, for simplicity, we assume that all D2D user groups $\mathcal{N}(t)$ are served in a TDMA fashion. Further improvement can be achieved by allowing parallel transmissions within multiple groups.}
\begin{equation}\label{eq:D2D_multicast_rate}
R_i^{\mathcal{N}} = \min_{k \in \mathcal{R}^{\mathcal{N}}(i)}\log\left( 1+\frac{P_d \lVert h_{ik} \rVert^2}{N_0}\right),
\end{equation} 
where $P_d$ is the device's transmit power constraint, and $h_{ik}$ is the channel response from user $i$ to user $k$. It should be noted that in each D2D transmission we assume that each user in $\mathcal{N}$, multicasts a message to a group of user. Thus, the rate is limited by the weakest receiver. 

In the downlink phase, the \ac{BS} multicasts coded messages containing all the remaining file fragments, such that, all of the users will be able to decode their requested content. 
\begin{figure}
    \includegraphics[width=0.8\columnwidth,keepaspectratio]{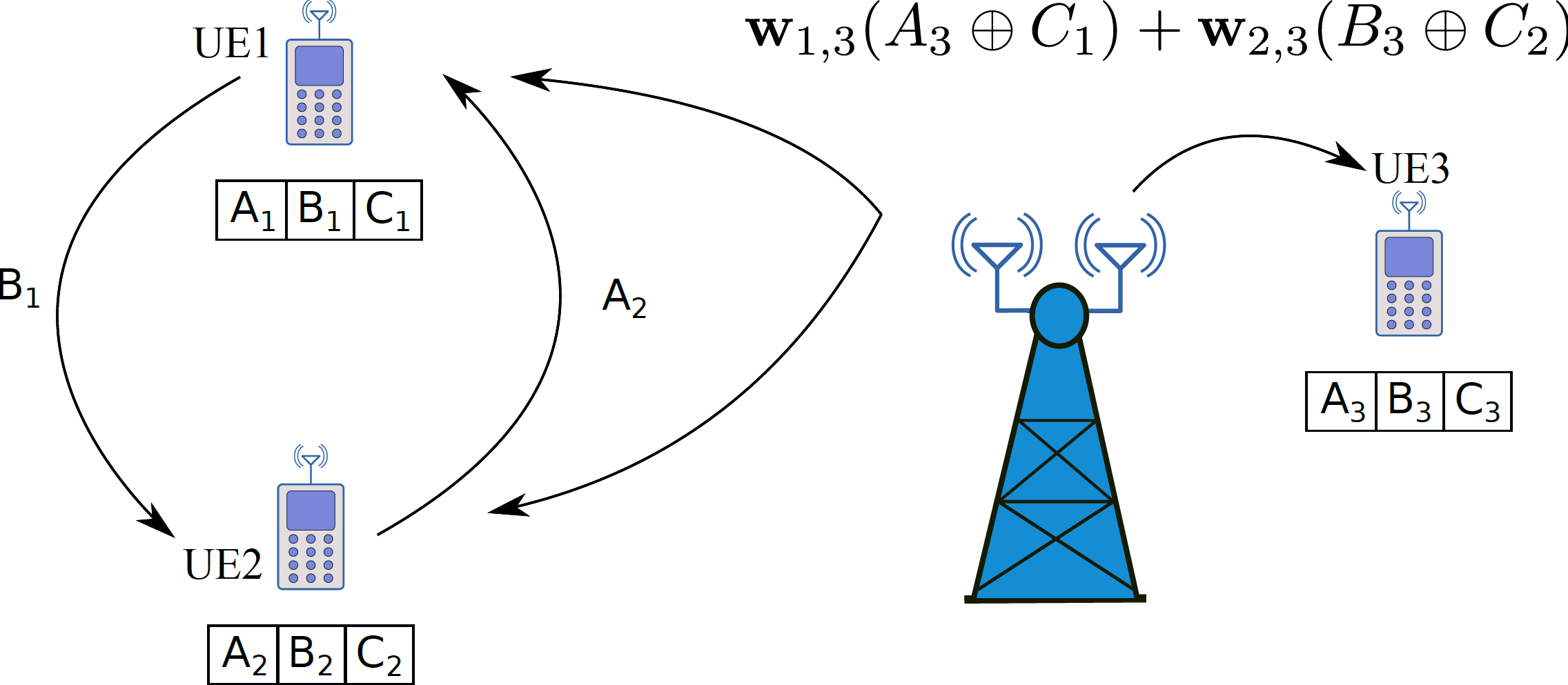}
    \caption{Example 1: D2D enabled downlink beamforming system model.}
\label{fig:sysmodel}
\end{figure}
The received downlink signal at user terminal $k = 1, \ldots, K$ is given by \begin{equation}
\label{eq:recv_signal}
    y_k = \M{h}_k\herm \sum_{ \mathcal{T} \subseteq \mathcal{S}} \mathbf{w}_{ \mathcal{T}}^{\mathcal{S}} \tilde{X}_{ \mathcal{T}}^{\mathcal{S}} + z_k
    \text{,}
\end{equation}
where $\tilde{X}_{ \mathcal{T}}^{\mathcal{S}}$ is the modulated version of the intended message $X_{ \mathcal{T}}^{\mathcal{S}}$ to be decoded by all the users in subset $ \mathcal{T}$ of set $\mathcal{S} \subseteq [1 : K] $, and $\mathbf{w}_{\mathcal{T}}^{\mathcal{S}}$ is the corresponding beamforming vector. The channel vector between the \ac{BS} and user $k$ is $\M{h}_k \in \mathcal{C}^L$, and the receiver noise is given by $z_k \sim \mathcal{N}(0, N_0)$. The \ac{CSIT} of all $K$ users is assumed to be perfectly known. 
The final achievable rate (per user) over the above-described two phases is given by 
\begin{equation}\label{eq:total_rate}
    R_{U}=\frac{F}{T_\text{D2D} + T_\text{DL}}
    \text{,}
\end{equation}
where $T_\text{D2D}$ and $T_\text{DL}$ denote the time used for the \ac{D2D} and \ac{DL} transmission sub-phases, respectively.

%% file: examples.tex

\section{D2D Aided Beamforming Explained: Examples}
\label{sec:examples}

In this section, we discuss the main concepts of the proposal via two examples. In the first example, we have a network of $3$ users, and in the second example, the number of users is increased to $4$.

\subsection{Example 1: $K=3$, $N=3$, $M=1$, and $L=2$}
In this example illustrated in Fig.~\ref{fig:sysmodel}, we have $K=3$ users and a library $\mathcal{W}=\{A,B,C\}$ of $N=3$ files, where each user has the cache size for storing just $M=1$ file. The base station is equipped with $L=2$ transmit antennas. To begin with, the cache content $Z_k$ at each user $k = 1,\ldots,K$ is 
\begin{equation}
    Z_1 = \{A_1, B_1, C_1\}, Z_2 = \{A_2, B_2, C_2\}, Z_3 = \{A_3, B_3, C_3\}\nonumber
\end{equation}
where we have assumed that each file is divided into three equal-sized sub-files. This follows the same cache placement as in~\cite{MaddahAli-2014}. In this example, we assume that users $1$ and $2$ are in close proximity, while user $3$ is far from them (see Fig. \ref{fig:sysmodel}). To describe the idea let us assume that users $1$, $2$, and $3$ request files $A$, $B$, and $C$, respectively. Now, the actual transmission strategy is split into two phases. In the first phase, which is called as the D2D sub-phase, users $1$ and $2$ are assumed to be using \ac{D2D} transmission to share their local cache content. Thus, the D2D sub-phase consists of a single D2D time slot with $\mathcal{N}=\{1,2\}$. It is evident that user $2$ would request $B_1$ from user $1$ and user $1$ would request $A_2$ from user $2$, and, since the \ac{D2D} transmission is assumed to be half duplex and requires TDMA, this single time slot constitutes of two D2D transmissions. The time required for the \ac{D2D} sub-phase is given by 
\begin{align} \nonumber
    T_\text{D2D} &= T\left(1 \rightarrow \mathcal{R}^{\mathcal{N}}(1)\right)+ T\left(2 \rightarrow \mathcal{R}^{\mathcal{N}}(2)\right)\\
    &= \frac{F / 3}{R^{\mathcal{N}}_1} + \frac{F / 3}{R^{\mathcal{N}}_2}
    \text{,} 
\end{align}
where $\mathcal{R}^{\mathcal{N}}(1)=\{2\}$, $\mathcal{R}^{\mathcal{N}}(2)=\{1\}$, and
\begin{align} \nonumber
R^{\mathcal{N}}_1=\log\left(1 + \frac{P_d \lVert h_{12} \rVert^2}{N_0}\right),
R^{\mathcal{N}}_2=\log\left(1 + \frac{P_d \lVert h_{21} \rVert^2}{N_0}\right).
\end{align}
Note that, in each transmission, $\frac{F}{3}$ fraction of the corresponding file is transmitted.

In the second (DL) sub-phase, the \ac{BS} multicasts the remaining content via coded messages. User $3$ was not active in the \ac{D2D} phase and still requires contents $C_1$ and $C_2$. However, users $1$ and $2$ only require $A_3$ and $B_3$, respectively. This content is XOR coded over two messages for user pairs $(1,3)$ and $(2,3)$. Namely, the messages are $X_{1,3} = A_3 \oplus C_1$ and $X_{2,3} = B_3 \oplus C_2$. 


Here, $X_{1,3}$ is a coded message, which would benefit users $1$ and $3$. Similarly, $X_{2,3}$ is a coded message intended for users $2$ and $3$. Thus, in order to deliver the correct coded message to each user, multicast beamformer vectors $\mathbf{w}_{1,3}$ and $\mathbf{w}_{2,3}$ are associated with messages $X_{1,3}$ and $X_{2,3}$, respectively. The downlink signal follows as 
    $\mathbf{x}_{DL}=\tilde{X}_{1,3}\mathbf{w}_{1,3}+\tilde{X}_{2,3}\mathbf{w}_{2,3}$, where $\tilde{X}_{1,3}$ and $\tilde{X}_{2,3}$ are the modulated messages (for more details see \cite{Tolli-Shariatpanahi-Kaleva-Khalaj-Arxiv18}).
Note that, here, user $3$ is assumed to use SIC receiver to decode both intended messages (interpreted as a multiple access channel (MAC)), while, users $1$ and $2$ only get served with a single message with the other seen as interference.  

Suppose now user $3$ can decode \emph{both} of its required messages $X_{1,3}$ and $X_{2,3}$ with the equal rate\footnote{Symmetric rate is imposed to minimize the time needed to receive both messages  $\tilde{X}_{1,3}$, and $\tilde{X}_{2,3}$.} 
\begin{equation}		
R^3_{MAC}=\min \left(\frac{1}{2} R^3_{Sum}, R^3_1, R^3_2\right),  
\end{equation} 
where the rate region corresponding to $\tilde{X}_{1,3}$, and $\tilde{X}_{2,3}$, is limited by
    $R^3_1 =\log\left(1+\frac{|\mathbf{h}_3^H \mathbf{w}_{1,3}|^2} {N_0}\right)$,
    $R^3_2=\log\left(1+\frac{|\mathbf{h}_3^H \mathbf{w}_{2,3}|^2} {N_0}\right)$
    and $R^3_{Sum}=\log\left(1+\frac{|\mathbf{h}_3^H \mathbf{w}_{1,3}|^2+|\mathbf{h}_3^H \mathbf{w}_{2,3}|^2} {N_0}\right)$.

Accordingly, the corresponding downlink beamformer design problem can be expressed as
\begin{equation}
\max_{\mathbf{w}_{2,3}, \mathbf{w}_{1,3}}  \min (R^3_\text{MAC}, R^1_1, R^2_1),
\end{equation}
where the rates of users $1$ and $2$ are given as
\begin{align}
    R^1_1&=\log\left(1+\frac{|\mathbf{h}_1^H \mathbf{w}_{1,3}|^2} {|\mathbf{h}_1^H \mathbf{w}_{2,3}|^2+N_0}\right) \\
    R^2_1&=\log\left(1+\frac{|\mathbf{h}_2^H \mathbf{w}_{2,3}|^2} {|\mathbf{h}_2^H \mathbf{w}_{1,3}|^2+N_0}\right).
\end{align}

Due to  D2D transmissions, the beamformer design problem is different as compared to~\cite{Tolli-Shariatpanahi-Kaleva-Khalaj-Arxiv18}. The partial file exchange in the D2D phase alleviates the interference conditions of the DL phase, thus, making the DL multicasting more efficient and less complex. 
On the other hand, the D2D transmission requires an orthogonal allocation in time domain. This introduces an inherent trade-off between the amount of resources allocated to the D2D and DL phases.

Finally, the corresponding symmetric rate maximization is given as
\begin{equation}
\begin{array}{rl}
	\label{prob:dlprob_ex1}
	\ds
	\underset{\substack{\M{w}_{i,j}, \gamma^k_l, r}}{\max} & 
	\ds
        r
	\\
	\mathrm{s.\ t.} 
    &  \ds r \leq \frac{1}{2} \log(1 + \gamma^3_1 + \gamma^3_2) \\
    &  \ds r \leq \log(1 + \gamma^3_1), \ r \leq \log(1 + \gamma^3_2) \\
    &  \ds r \leq \log(1 + \gamma^1_1), \ r \leq \log(1 + \gamma^2_1) \\
    & \ds \gamma^1_1 \leq \frac{|\M{h}_1\herm\M{w}_{1,3}|^2}{|\M{h}_1\herm\M{w}_{2,3}|^2 + N_0}, 
          \gamma^2_1 \leq \frac{|\M{h}_2\herm\M{w}_{2,3}|^2}{|\M{h}_2\herm\M{w}_{1,3}|^2 + N_0} 
    \\
    & \ds \gamma^3_1 \leq \frac{|\M{h}_3\herm\M{w}_{1,3}|^2}{N_0}, 
          \gamma^3_2 \leq \frac{|\M{h}_3\herm\M{w}_{2,3}|^2}{N_0} \\
	& \ds 
        \|\M{w}_{1,3}\|^2 + \|\M{w}_{2,3}\|^2 \leq \text{SNR}
		\text{.}
\end{array}
\end{equation}
The rate constraints can be written as convex second-order cone constraints as shown in~\cite{Tolli-Shariatpanahi-Kaleva-Khalaj-Arxiv18}. However, the \ac{SINR} constraints are non-convex and require an iterative solution. A \ac{SCA} solution for the \ac{SINR} constraints can be found, e.g., in~\cite{Tolli-Shariatpanahi-Kaleva-Khalaj-Arxiv18}. Please notice that, here due to D2D transmission in the the first phase we have only two beamformer vectors ($\M{w}_{1,3}$ and $\M{w}_{2,3}$), which means that we can dedicate more power to our intended signals ($X_{1,3}$ and $X_{2,3}$)  compared to~\cite{Tolli-Shariatpanahi-Kaleva-Khalaj-Arxiv18}. The time required for the \ac{DL} phase is given by 
\begin{equation}
    T_\text{DL}= \frac{F/3}{r} = \frac{F/3}{\max_{\mathbf{w}_{2,3}, \mathbf{w}_{1,3}}  \min (R^3_\text{MAC}, R^1_1, R^2_1)}
    \text{,}
\end{equation}
Note that, also in this phase, all users are served with coded messages of size $\frac{F}{3}$ bits, which are multiplexed with the help of the beamforming vectors. Finally, the achievable rate over the D2D and DL phases is given in~\eqref{eq:total_rate}.

\subsection{Example 2: $K=4$, $N=4$, $M=2$, and $L=2$}
In this example, we have $K=4$ users and a library $\mathcal{W}=\{A,B,C,D\}$ of $N=4$ files, where each user has a cache for storing $M=2$ files. Also, the base station is equipped with $L=2$ transmit antennas. Following the same placement as in \cite{MaddahAli-2014} (from now on we note $t=KM/N$), each file is split into $\binom{K}{t}=\binom{4}{2}=6$ subfiles, as follows
\begin{align} \nonumber
A&=\{A_{1,2}, A_{1,3}, A_{1,4}, A_{2,3}, A_{2,4}, A_{3,4}\}, \\ \nonumber
B&=\{B_{1,2}, B_{1,3}, B_{1,4}, B_{2,3}, B_{2,4}, B_{3,4}\}, \\ \nonumber
C&=\{C_{1,2}, C_{1,3}, C_{1,4}, C_{2,3}, C_{2,4}, C_{3,4}\}, \\ \nonumber
D&=\{D_{1,2}, D_{1,3}, D_{1,4}, D_{2,3}, D_{2,4}, D_{3,4}\}.
\end{align}
Each file $W_{\tau}$ is cached at user $k$ if $k \in \tau$. Let us assume that users $1-4$ request files $A-D$, respectively.

In this example, we suppose that users $1$, $2$, and $3$ are close to each other, while user $4$ is far from them. 
Then, the \ac{D2D} sub-phase consists of exchanging information between the first three users locally (collected in $\mathcal{N}=\{1,2,3\}$) in  three orthogonal D2D transmissions. More specifically, each subfile is divided into $t=KM/N=2$ parts which are discriminated by their superscript indices.
Then, in the first D2D transmission of length $T\left(1\rightarrow \mathcal{R}^{\mathcal{N}}(1)\right)$ seconds, user $1$ multicasts $X_1= B^{1}_{1,3} \oplus C^{1}_{1,2}$ to $\mathcal{R}^{\mathcal{N}}(1)=\{2,3\}$. In the second D2D transmission, user $2$ transmits $X_2= A^{1}_{2,3} \oplus C^{2}_{1,2}$ to $\mathcal{R}^{\mathcal{N}}(2)=\{1,3\}$, which will take $T\left(2\rightarrow \mathcal{R}^{\mathcal{N}}(2)\right)$ seconds. Finally, in the third D2D transmission of length $T\left(3\rightarrow \mathcal{R}^{\mathcal{N}}(3)\right)$ seconds, user $3$ transmits $X_3= A^{2}_{2,3} \oplus B^{2}_{1,3}$ to $\mathcal{R}^{\mathcal{N}}(3)=\{1,2\}$. 
These transmissions require the total time of
\begin{equation}
\begin{array}{rl}
    \ds T_{D2D} = & \ds T\left(1\rightarrow \mathcal{R}^{\mathcal{N}}(1)\right)+T\left(2\rightarrow \mathcal{R}^{\mathcal{N}}(2)\right)+
    \\ & \ds
    T\left(3\rightarrow \mathcal{R}^{\mathcal{N}}(3)\right)
\end{array}
\end{equation}
in which
\begin{math}
T\left(i\rightarrow \mathcal{R}^{\mathcal{N}}(i)\right) = \frac{F/12}{R^{\mathcal{N}}_i}, \quad i= {1,2,3}
\end{math}
and $R^{\mathcal{N}}_i, i={1,2,3}$ are determined by~\eqref{eq:D2D_multicast_rate}.
Then, in the DL sub-phase, the BS transmits the remaining messages
\begin{equation}
\mathbf{x}_{DL}=\tilde{X}_{1,2,4} \mathbf{w}_{1,2,4}+\tilde{X}_{1,3,4} \mathbf{w}_{1,3,4}+\tilde{X}_{2,3,4} \mathbf{w}_{2,3,4},
\end{equation}
where $\tilde{X}_{1,2,4}=A_{2,4} \oplus B_{1,4} \oplus D_{1,2}$, $\tilde{X}_{1,3,4}=A_{3,4} \oplus C_{1,4} \oplus D_{1,3}$, and $\tilde{X}_{2,3,4}=B_{3,4} \oplus C_{2,4} \oplus D_{2,3}$. At the end of this sub-phase, user $1$ is interested in decoding $\{X_{1,2,4}$, $X_{1,3,4}\}$, user $2$ is interested in decoding $\{X_{1,2,4}$, $X_{2,3,4}\}$, user $3$ is interested in decoding $\{X_{1,3,4}$, $X_{2,3,4}\}$, and finally, user $4$ is interested in decoding all the three terms $\{X_{1,2,4}, X_{1,3,4}$, $X_{2,3,4}\}$. Thus, from the perspective of users $1$, $2$, and $3$, we have a MAC channel with two useful terms and one interference term. However, from the perspective of the user $4$, we have a MAC channel with three useful terms. Thus, for users $1$, $2$, and $3$ we have MAC rate region
\begin{equation}		
R^k_\text{MAC}=\min (R^k_\text{sum}, 2 R^k_1, 2 R^k_2), \quad \quad k=1,2,3.
\end{equation} 
For example, for $k=1$, we have
    $R^1_1 =\log\left(1+\frac{|\M{h}_1\herm\M{w}_{1,2,4}|^2}{|\M{h}_1\herm\M{w}_{2,3,4}|^2 + N_0}\right)$, 
    $R^1_2 =\log\left(1+\frac{|\M{h}_1\herm\M{w}_{1,3,4}|^2}{|\M{h}_1\herm\M{w}_{2,3,4}|^2 + N_0}\right)$ and 
    $R^1_\text{sum}= \log \left(1+\frac{|\M{h}_1\herm\M{w}_{1,2,4}|^2+|\M{h}_1\herm\M{w}_{1,3,4}|^2}{|\M{h}_1\herm\M{w}_{2,3,4}|^2 + N_0}\right)$.

In order to derive the fourth user's 3-stream rate region, we face a MAC with three messages. Thus, we have $7$ MAC region inequalities, which will result in $R^4_\text{MAC}$ (the details are omitted here due to lack of space. For details refer to \cite{Toll1806:Multicast,Tolli-Shariatpanahi-Kaleva-Khalaj-Arxiv18}). When all the MAC inequalities for all the users are gathered together we can derive the common multicast rate, which is shown in the corresponding downlink beamformer design problem as follows
\begin{equation}
\begin{array}{l}
	\label{prob:dlprob_ex2}
		\underset{\substack{\M{w}_{i,j,l}, \gamma^k_m, r }}{\max}  
	
        r
	\\
	 \mathrm{subject  \ to} \\ 
       r \leq \frac{1}{2}\log(1 + \gamma^k_1 + \gamma^k_2),\ k=1,2,3 \\ 
       r \leq \log(1 + \gamma^k_m),\ k=1,2,3, m = 1,2 \\
      r \leq \frac{1}{3}\log(1 + \gamma^4_1 + \gamma^4_2+\gamma^4_3),  \ 
       r \leq \frac{1}{2}\log(1 + \gamma^4_1 + \gamma^4_2) \\ 
      r \leq \frac{1}{2}\log(1 + \gamma^4_1 + \gamma^4_3), \ 
      r \leq \frac{1}{2}\log(1 + \gamma^4_2 + \gamma^4_3) \\ 
      \ds r \leq \log(1 + \gamma^4_m),\ m = 1,2,3 \\
     \ds \gamma^1_1 \leq \frac{|\M{h}_1\herm\M{w}_{1,2,4}|^2}{|\M{h}_1\herm\M{w}_{2,3,4}|^2 + N_0}, 
          \gamma^1_2 \leq \frac{|\M{h}_1\herm\M{w}_{1,3,4}|^2}{|\M{h}_1\herm\M{w}_{2,3,4}|^2 + N_0} 
    \\
     \ds \gamma^2_1 \leq \frac{|\M{h}_2\herm\M{w}_{1,2,4}|^2}{|\M{h}_2\herm\M{w}_{1,3,4}|^2 + N_0}, 
          \gamma^2_2 \leq \frac{|\M{h}_2\herm\M{w}_{2,3,4}|^2}{|\M{h}_2\herm\M{w}_{1,3,4}|^2 + N_0} 
    \\
     \ds \gamma^3_1 \leq \frac{|\M{h}_3\herm\M{w}_{1,3,4}|^2}{|\M{h}_3\herm\M{w}_{1,2,4}|^2 + N_0}, 
          \gamma^3_2 \leq \frac{|\M{h}_3\herm\M{w}_{2,3,4}|^2}{|\M{h}_3\herm\M{w}_{1,2,4}|^2 + N_0} 
    \\
     \ds \gamma^4_1 \leq |\M{h}_4\herm\M{w}_{1,2,4}|^2/N_0, 
          \gamma^4_2 \leq |\M{h}_4\herm\M{w}_{1,3,4}|^2/N_0 \\
     \ds
          \gamma^4_3 \leq |\M{h}_4\herm\M{w}_{2,3,4}|^2/N_0 \\
	 \ds 
        \|\M{w}_{1,2,4}\|^2 + \|\M{w}_{1,3,4}\|^2 + \|\M{w}_{2,3,4}\|^2 \leq \text{SNR}
		\text{.}
\end{array}
\end{equation}
Finally, the delivery time of the DL sub-phase is
    $T_\text{DL} = \frac{F/6}{r}$.

It should be noted that, compared to the solution proposed in \cite{Tolli-Shariatpanahi-Kaleva-Khalaj-Arxiv18}, we have one term removed from the downlink transmission, i.e., $\tilde{X}_{1,2,3}\mathbf{w}_{1,2,3}$. This term is already taken care of in the D2D phase,  which in turn enhances the performance of the downlink phase. 

%% file: general.tex
\section{D2D Aided Beamforming: The General Case}
\label{sec:general}

In this section,  we formulate and analyze the proposed scheme in the general setting. 
The cache content placement phase is identical to the one proposed in \cite{MaddahAli-2014}. In general, in each data transmission, $\min(t+L,K)$ users can be served simultaneously \cite{Tolli-Shariatpanahi-Kaleva-Khalaj-Arxiv18}. Thus, when $t+L < K$, $\binom{K}{t+L}$ transmission phases are required in total. Unlike in~\cite{Tolli-Shariatpanahi-Kaleva-Khalaj-Arxiv18}, here, the data delivery is split into D2D and DL sub-phases. 

To examine the optimal D2D sub-phase user allocation, we need to perform exhaustive search of the D2D subsets. There are, in total, $\binom{t+L}{t+1}$ different user subsets (of size $t+1$) among $t+L$ number of users in each transmission phase.  
Thus, the exhaustive search would require $2^{\binom{t+L}{t+1}}$ evaluations of~\eqref{eq:total_rate}. In each of these evaluations, all the beamformers must be solved and total rate computed. Then, the highest one should be chosen.   To simplify the notation, we consider an indication function $I_{D2D}(\mathcal{T})$, which specifies whether the corresponding subset has been allocated for D2D transmission. 
We define $C(K,t,L)=\frac{F}{\binom{K}{t}\binom{K-(t+1)}{L-1}}$ as the size of the transmitted file fragment~\cite{Tolli-Shariatpanahi-Kaleva-Khalaj-Arxiv18}.

{\subsection{Total delivery time $T_{\mathrm{D2D}}+T_{\mathrm{DL}}$}\label{sec:delivery_time_general}
Now, for a given D2D mode allocation, the D2D delivery time is given as 
\begin{align}
    &T_{\mathrm{D2D}}  = \sum_{\mathcal{T} \subseteq \overline{\Omega^\mathcal{S}}}\sum_{k \in \mathcal{T}}\frac{C(K,t,L)/t}{\mathcal{R}^{\mathcal{N}}_{k}}\label{Eq:pd2d},
\end{align}
where}
\begin{math} 
    \overline{\Omega^\mathcal{S}} := \{\mathcal{T} \subseteq \mathcal{S}, |\mathcal{T}|=t+1, I_{\mathrm{D2D}}(\mathcal{T})=1\}
\end{math} and $\mathcal{R}^{\mathcal{N}}_{k}$ is from~\eqref{eq:D2D_multicast_rate}.
Since in each D2D subset each file fragment is transmitted by $t$ users, we further divide each file fragment in to $t$ sub-packets so that we can transmit a distinct sub-packet by each user (see Example 2).

The beamformers for the DL phase are solved using the SCA approach from~\cite{Tolli-Shariatpanahi-Kaleva-Khalaj-Arxiv18}. The main difference, in contrast to~\cite{Tolli-Shariatpanahi-Kaleva-Khalaj-Arxiv18}, is that we should not consider all the $t+1$ subsets. Here, only those subsets $\mathcal{T}$ for which $I_{\mathrm{D2D}}(\mathcal{T})=0$ should be involved in the DL phase. This will reduce the interference between parallel streams significantly. The DL sub-phase throughput is given by
	\begin{align}\label{Eq:MulticastRate} \nonumber
&R_C\left(\mathcal{S},\{\mathbf{w}_\mathcal{T}^\mathcal{S}, \mathcal{T} \subseteq \mathcal{S}, |\mathcal{T}|=t+1, I_{D2D}(\mathcal{T})=0\}\right)= \\
& \min_{k \in \mathcal{S}} R^k_{MAC}\left(\mathcal{S},\{\mathbf{w}_\mathcal{T}^\mathcal{S}, \mathcal{T} \subseteq \mathcal{S}, 
I_{\mathrm{D2D}}(\mathcal{T})=0\}\right)
	\end{align}
where
\begin{align}\label{Eq:MAC_general}
	&R^k_{MAC}\left(\mathcal{S},\{\mathbf{w}_\mathcal{T}^\mathcal{S}, \mathcal{T} \subseteq \mathcal{S}, 
	I_{\mathrm{D2D}}(\mathcal{T})=0\}\right) \nonumber \\ 
	= &\min_{\mathcal{B} \subseteq \Omega_k^\mathcal{S}} \left[\frac{1}{|\mathcal{B}|}\log\left(1+\frac{\sum_{\mathcal{T} \in \mathcal{B}} |\mathbf{h}_k\herm \mathbf{w}_\mathcal{T}^\mathcal{S}|^2}{N_0+\sum_{\mathcal{T} \in \Omega_\mathcal{S} \backslash \Omega_k^\mathcal{S}} |\mathbf{h}_k\herm \mathbf{w}_\mathcal{T}^\mathcal{S}|^2}\right) \right]
		\end{align}
		where 
		\begin{align}
		     &\Omega^\mathcal{S} := \{\mathcal{T} \subseteq \mathcal{S}, |\mathcal{T}|=t+1, I_{\mathrm{D2D}}(\mathcal{T})=0\} \\
		     &\Omega_k^\mathcal{S} := \{\mathcal{T} \subseteq \mathcal{S}, |\mathcal{T}|=t+1, I_{\mathrm{D2D}}(\mathcal{T})=0 \ | \ k \in \mathcal{T}\}.
		\end{align}

After computing the rate for DL sub-phase the $T_\mathrm{DL}$ is computed as $T_\mathrm{DL}=\frac{C(K,t,L)}{R_C}$, then the achievable symmetric rate per user is computed using \eqref{eq:total_rate}. 
For a large number of users and transmit antennas, solving~\eqref{Eq:MulticastRate} requires a considerable amount of computation, due to the iterative convex approximation for each subset evaluation~\cite{Tolli-Shariatpanahi-Kaleva-Khalaj-Arxiv18}. 
In the following, we provide a low complexity heuristic solution for the proposed mode assessment problem.

\subsection{Heuristic D2D mode selection with low complexity}
In order to decrease the computational load of evaluating ${T}_{\mathrm{D2D}}$ and ${T}_{\mathrm{DL}}$ for different D2D mode allocations, we provide a throughput approximation for the D2D mode allocations without having to rely on the general SCA solution for the DL beamformer design. 
The D2D transmissions occur in orthogonal time slots. The accumulated D2D phase duration is denoted by $T_{\mathrm{D2D}}$. Each successful D2D exchange reduces the remaining number of file fragments to be transmitted by the BS. Thus, there are fewer multicast messages and corresponding beamforming vectors $\mathbf{w}_{\mathcal{T}}^{\mathcal{S}}$ in the DL optimization problem. This allows more efficient (less restricted) multicast beamformer design, which results in reduced DL phase duration $T_{\mathrm{DL}}$. The D2D mode selection 
is iteratively carried out as long as the following condition holds: 
%
%
\begin{equation} 
\label{eq:therishold}
  \frac{\hat{T}^{i}_{\mathrm{DL}}}{N_{\mathrm{F}}-(t+1)(i-1)} \geq \hat{T}_{\mathrm{D2D}}^i, i\in\big[1, \binom{t+L}{t+1}\big],
\end{equation}
where $N_{\mathrm{F}} = (t+1)\binom{t+L}{t+1}$ is the total number of file fragments that should be delivered to all the users so that they can decode their intended files. Moreover, $\hat{T}^{i}_{\mathrm{DL}}$ and $\hat{T}_{\mathrm{D2D}}^i$ are the coarse approximated delivery times in the $i^\text{th}$ iteration.
In~\eqref{eq:therishold}, we check if any D2D user subset will reduce the DL duration $T_{\mathrm{DL}}$ more than the duration of the corresponding D2D transmission. If a specific subset $\mathcal{T}$ in iteration $i$ satisfies~\eqref{eq:therishold}, then the D2D transmission for this subset is done following the approach proposed in \cite{Ji2016}. 

In each D2D time slot, 
$t+1$ fragments of files are delivered by $t+1$ orthogonal D2D transmissions. On the other hand, in the DL sub-phase, all the remaining fragments ($N_{\mathrm{F}}-(t+1)(i-1)$) are delivered simultaneously. Thus, in~\eqref{eq:therishold}, the average delivery time for one fragment in the D2D and DL phases are compared. 
 In each iteration, we choose a subset for D2D candidate, i.e., the subset which provides the highest rate. If, at any specific iteration,~\eqref{eq:therishold} does not hold, using more D2D transmissions will not improve the overall rate and the iterative process is terminated. Therefore, at most $\binom{t+L}{t+1}$ iterations are required compared to $2^{\binom{t+L}{t+1}}$ needed for the exhaustive search.



The D2D delivery time is coarsely approximated as 
\begin{align}
    &\hat{T}_{\mathrm{D2D}}^i  = \frac{C(K,t,L)/t}{\hat{R}^{i}_{\mathrm{D2D}}}, \
    \hat{R}^{i}_{\mathrm{D2D}}=\max_{\mathcal{T} \subseteq \Omega^\mathcal{S}}\hat{R}^{i}_{\mathcal{T}}, \nonumber \\
    &{\hat{R}^{i}_{\mathcal{T}}=\frac{1}{(t+1)} \sum_{k \in \mathcal{T}}\min_{j \in \mathcal{R}^{\mathcal{N}}(k)}\log\left( 1+\frac{P_d \lVert h_{kj} \rVert^2}{N_0}\right)}\label{Eq:apd2d},
\end{align}
  
  Since, in each D2D transmission (e.g., user $i$'s transmission in Fig. \ref{fig:timeDiv}), $1/t$ part of each fragment is delivered, $\hat{T}_{\mathrm{D2D}}^i$ is considered as  $\frac{C(K,t,L)/t}{\hat{R}^{i}_{\mathrm{D2D}}}$ to scale the delivery time. Here, the approximated D2D rate for each subset is simply defined as the average rate of the users in that subset. In each D2D subset there are $t+1$ number of users which transmit a data useful for $t$ number of other users, thus in total there are $(t+1)$ terms in \eqref{Eq:apd2d}. 
 Note that, for each iteration $i$, we only consider those subsets that have not already been allocated for D2D.
 
 The DL delivery time is coarsely approximated as
\begin{align}
    &\hat{T}^{i}_{\mathrm{DL}} = \frac{C(K,t,L)}{\hat{R}^{i}_{\mathrm{DL}}}, \quad
    \hat{R}^{i}_{\mathrm{DL}} = {\min_{k\in [\mathcal{S}]}}\hat{R}^{i}_{k}, \nonumber \\
    &\hat{R}^{i}_{k} = \min_{\mathcal{B} \subseteq \Omega_k^\mathcal{S}} \left[\frac{1}{|\mathcal{B}|}\log\left(1+\frac{\text{SNR}}{|\Omega^\mathcal{S}|N_0}\sum_{\mathcal{T} \in \mathcal{B}} \frac{1}{|\mathcal{T}|} \sum_{j \in \mathcal{T}} \lVert \mathbf{h}_j \rVert^2  \right) \right],
\end{align}
where
$\hat{R}^{i}_{k}$ is the approximated rate of user $k$ considering that ($i-1$) subsets had been chosen for D2D transmission in the previous iterations. 
Here, for simplicity, we omit the interference among parallel multicast streams and consider equal power loading over all the remaining subsets of users ($\frac{\text{SNR}}{|\Omega^\mathcal{S}|N_0}$). 
In general, beamformers $\mathbf{w}_\mathcal{T}^\mathcal{S}$ should be designed in such a way that all the users in subset $\mathcal{T}$ can decode the message $\hat{X}_\mathcal{T}^\mathcal{S}$. For the heuristic mode selection process, however, we simply use the average channel gain assuming matched filter beamforming ($\frac{1}{|\mathcal{T}|} \sum_{j \in \mathcal{T}} \lVert \mathbf{h}_j \rVert^2$) to coarsely indicate the multicast beamforming potential for a given subset.
Once the users for D2D mode transmission are found based on~\eqref{eq:therishold}, the final delivery time and the rate are computed as described in Section~\ref{sec:delivery_time_general}.

%% file: simres.tex
\section{Numerical Examples}
\label{sec:simres}

%
%
For better understanding the network level impact of D2D transmission, we have simulated general scenarios for $K=3$ and $K=4$ users cases. In our scenarios, users are scattered in a circular area with radius of $R=100$ meters. Moreover, in order to see the effect of D2D transmission in different situations we control the maximum user separation. Thus, we have considered another circle inside the cell area which can be located anywhere inside the cell and users are scattered inside this smaller circle. In this manner the maximum distance between two users is $2r$ (r is the radius of smaller circle) but the users distance to BS is any number between 0 to $R$ ($R$ is the radius of cell). Thus, by changing $r$ we can control the maximum users’ separation in D2D mode which helps us investigate the beneficial users distance in D2D mode. The channel coefficients of the users are as
\begin{math}
    h_{j,k}=(\frac{1}{d_{j,k}})^{\frac{n}{2}}G,\ \text{for} \ j= 1, \dots, K \ \text{and} \ j \neq k,
    k \in \{1, \dots, K\} \cup \{BS\},
\end{math}
where $G$ is a complex Gaussian variable with zero mean and unit variance, $n$ is the path loss exponent ($3$ for DL and $2$ for D2D), $d$ is the users distance from the transmitter (BS in DL and user in D2D). 

\begin{figure}
    \centering 
    \includegraphics[width=\columnwidth,keepaspectratio]{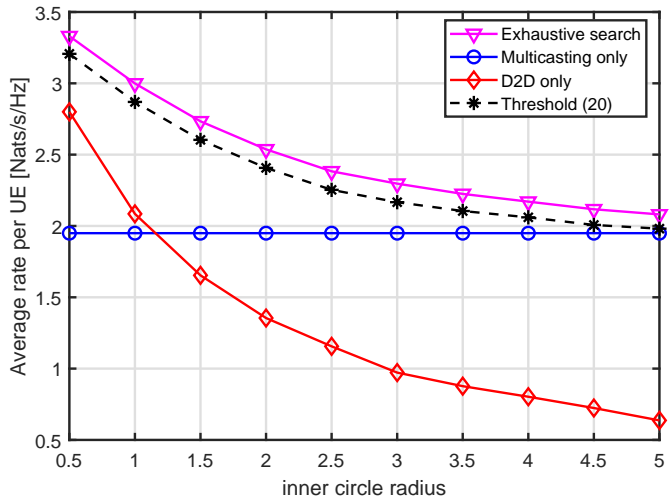}
    \caption{Per user rate vs. small circle radius $r$ for $K=3$ and $t=1$.}
\label{fig:NA3}
\end{figure}

Transmit powers at users for D2D transmission and at the BS for DL multicast beamforming are adjusted in a way that, the received SNR is $0$ dB at $10$ meter distance from another user, and at the cell edge ($100$m distance between a user and BS), respectively. 
Fig.~\ref{fig:NA3} shows the per user rate for $K=3$ case (Example $1$) as a function of inner circle radius.
Fig.~\ref{fig:NA3} demonstrates that, when users are close to each other, we have a significant gain from using a combination of multicasting and D2D transmissions. However, when the maximum distance between users start to increase the rate of 'D2D only' transmission decreases drastically. 
The reasonable range for using D2D transmission in particular scenarios is between $r= 0 \ \text{and} \ 5$m ($10$m maximum distance), while this distance changes by path loss exponent, D2D and DL available power, $t$, etc. 

\begin{figure}
    \centering 
    \includegraphics[width=\columnwidth,keepaspectratio]{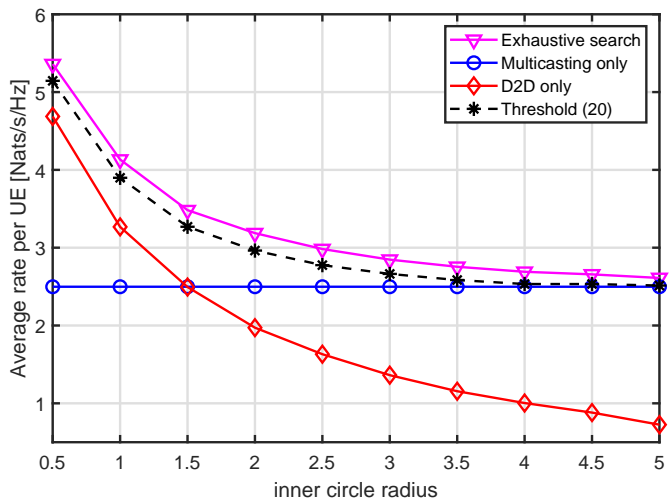}
    \caption{Per user rate vs. small circle radius $r$ for $K=4$ and $t=2$.}
\label{fig:NA4}
\end{figure}
Fig.~\ref{fig:NA4} shows the per user rate versus inner circle radius  for $K=4,\ t=2,\ \text{and} \ L=2$ (Example $2$). For a higher number of users the gain from using D2D transmission among nearby users is clearly larger than in~\ref{fig:NA3}. However, the gain of D2D transmission decreases more rapidly compared to the case $K=3$. Since $t=2$, we need more users to be closer to each other in order to be able to perform the D2D transmission in an efficient manner. 

It is worth to mention that, using the heuristic D2D mode selection criteria defined in Section \ref{sec:general} results in minimal loss in per user rate, with a greatly reduced  complexity, as compared to the exhaustive search. 

%% file: conclusions.tex

\section{Conclusions}
\label{sec:conclusions}
A novel delivery scheme optimized for finite \ac{SNR} region was proposed, where the multicast beamforming of file fragments is complemented by allowing direct \ac{D2D} exchange of local cache content. The benefits of partial D2D offloading of multicast delivery of coded caching content were investigated. Two simple example scenarios were assessed in detail and a generalized formulation was also provided. 
Moreover, a heuristic low complexity mode selection scheme was proposed with comparable performance to the optimal exhaustive search. In the future work, we will provide a detailed complexity analysis of the proposed scheme, which will formalize the gains related to the computational complexity.
